\documentclass[preprint,showpacs,preprintnumbers,amsmath,amssymb,prb]{revtex4}

\usepackage{graphicx}
\usepackage{dcolumn}
\usepackage{bm}

\begin{document}

\preprint{APS/QED-TMartin-APL-v2}

\title{Enhanced Zeeman splitting in Ga$_{0.25}$In$_{0.75}$As quantum point contacts}

\author{T. P. Martin}
\email{tmartin@phys.unsw.edu.au}
\author{A. Szorkovszky}%
\author{A. P. Micolich}%
\author{A. R. Hamilton}%
\affiliation{School of Physics, University of New South Wales,
Sydney NSW 2052, Australia}

\author{C. A. Marlow}%
\author{H. Linke}%
\author{R. P. Taylor}
\altaffiliation{Department of Physics and Astronomy, University of
Canterbury, Christchurch 8140, New Zealand}
\affiliation{Department
of Physics, University of Oregon, Eugene OR 97403, USA}

\author{L. Samuelson}%
 \affiliation{Division of Solid State Physics, Lund University, Box 118, S-221 00 Lund, Sweden}

\date{\today}

\begin{abstract}
The strength of the Zeeman splitting induced by an applied magnetic
field is an important factor for the realization of spin-resolved
transport in mesoscopic devices.  We measure the Zeeman splitting
for a quantum point contact etched into a Ga$_{0.25}$In$_{0.75}$As
quantum well, with the field oriented parallel to the transport
direction. We observe an enhancement of the Land\'e \emph{g}-factor
from $|g^{*}|=3.8\pm0.2$ for the third subband to
$|g^{*}|=5.8\pm0.6$ for the first subband, six times larger than in
GaAs. We report subband spacings in excess of 10 meV, which
facilitates quantum transport at higher temperatures.
\end{abstract}

\maketitle

The control of a charge carrier's spin is typically achieved using
an applied magnetic field, and provides an extra degree of freedom
that can be utilized for device functionalities such as
spintronics, quantum information,
etc.~\cite{wolf01,elzerman04,craig04,tombros06} In semiconductor
devices, the field $B$ breaks the degeneracy of the two spin states
via the Zeeman effect, resulting in spin-polarized transport, where
a particular spin orientation dominates the electrical conductance
of the device.~\cite{wolf01,elzerman04} The Zeeman spin-splitting is
given by $\Delta E_{z}=g^{*}\mu_{B}B$, where $g^{*}$ is the
effective Land\'e \emph{g}-factor and $\mu_{B}$ is the Bohr
magneton. Since the strength of the spin-splitting is governed by
$g^{*}$, narrow band-gap materials with a large $g$-factor such as
GaInAs or InAs are highly desirable in the quest to develop
electronic devices that require only the smallest magnetic fields to
achieve spin-functional operations.

The small band-gap in the Ga$_{x}$In$_{1-x}$As material system makes
a large $g$-factor possible due to mixing of the conduction band
electron states with valence band states. For example, in GaInAs
quantum wells, $g$-factors ranging from 2.9 to 4.4 have been
measured~\cite{vehse86,dobers89,savelev96,schapers07} -- an order of
magnitude larger than in equivalent GaAs quantum
wells.~\cite{dobers88,dobers89} It is thus interesting to know how
the $g$-factor in lower-dimensional structures will behave, because
the additional confinement alters the coupling between the
conduction and valence bands. In both $n$-type and $p$-type GaAs quantum point contacts (QPCs),
the confinement of electrons to a quasi-one-dimensional (quasi-1D) system can
lead to an enhancement of $|g^{*}|$ by as much as a factor of two
over its 2D value, with the enhancement increasing as the 1D
confinement is strengthened.~\cite{patel91a,thomas96,daneshvar97,danneau06}  
Unexpectedly, the only measurement to date of the $g$-factor in a GaInAs
QPC~\cite{schapers07} gave $|g^{*}|\approx4$ in the 1D limit,
showing no clear enhancement over the value of $|g^{*}|$
obtained in the 2D reservoirs adjacent to the QPC.~\cite{schapers07}

In this letter, we use parallel field measurements (oriented in the
plane of the quantum well along the QPC) to study how the strength
of the 1D confinement affects the spin-splitting in an etched
Ga$_{0.25}$In$_{0.75}$As QPC. In contrast to the measurements
obtained using perpendicular fields by Sch\"apers \emph{et
al.},~\cite{schapers07} we observe a clear enhancement of the
$g$-factor in the 1D limit of our device. Our result is consistent
with the 1D enhancement observed in GaAs QPCs, which was also
measured using parallel field techniques.~\cite{patel91a,thomas96,daneshvar97,danneau06}
The key difference between our measurements and those of
Reference~8 is the field direction, and we use this to
explain why 1D $g$-factor enhancement was not observed in their
study.

Our QPC is etched into a GaInAs/InP modulation-doped
heterostructure, where a 2D electron gas (2DEG) is confined to a 9
nm thick Ga$_{0.25}$In$_{0.75}$As quantum well.~\cite{ramvall97,martin08b} 
The QPC investigated is $\sim$160 nm long and $\sim$120 nm wide, and
fabricated on a Hall bar mesa featuring NiGeAu Ohmic contacts. A
Ti/Au top-gate, deposited uniformly over the mesa, is used to tune
the Fermi energy $E_{F}$ and thus change the number of occupied
subbands $n$ in the QPC. All measurements were performed in a
$^{4}$He cryostat with a base temperature of 1.3~K. Standard
four-probe and lock-in amplification techniques were used to measure
the differential conductance $G=dI/dV$ through the QPC at a
frequency of 17 Hz and constant excitation voltage of 100 $\mu$V
(300 $\mu$V for source-drain biasing). The 2D carrier density and
mobility were 5.6--6.8$\times10^{11}$ cm$^{-2}$ and
1.3--2.2$\times10^{5}$ cm$^{2}$/Vs respectively within the range of
applied gate voltages presented here ($-6.0$ to 1.5 V). Our 2D
density and mobility, gallium fraction $x=0.25$, and quantum well
width are very similar to the values reported in
Ref.~8.

\begin{figure}
\includegraphics[width=8.5cm]{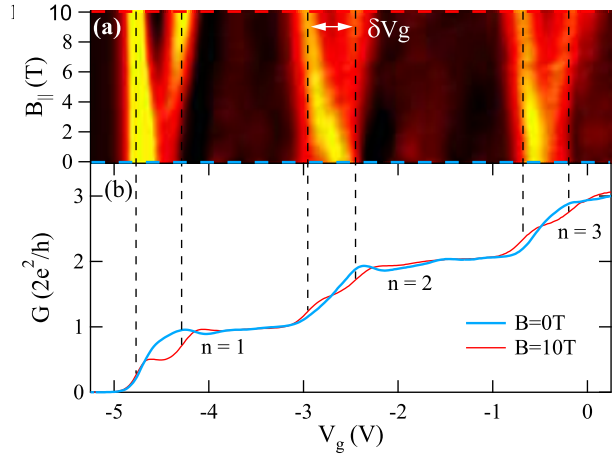}
\caption{\label{fig:1} (Color online) Zeeman splitting for the first
three subbands of the QPC. (a) Transconductance $dG/dV_{g}$ as a
function of gate voltage $V_{g}$ (horizontal axis) and magnetic
field $B_{\parallel}$ (vertical axis). Regions of yellow and red
correspond to a large amplitude of $dG/dV_{g}$, indicating the
locations of subband transitions. (b) Conductance $G$ vs gate
voltage $V_{g}$ for $B_{\parallel}=0$ T (thick blue line) and 10 T
(thin red line). Dashed lines in (a) indicate the region of the plot
corresponding to the colored traces in (b).}
\end{figure}

The blue line in Figure~\ref{fig:1}(b) shows the conductance $G$ of
the QPC at $B=0$, demonstrating clear plateaux as a function of gate
voltage $V_{g}$. The plateaux are a consequence of the 1D
confinement in the QPC, where each occupied subband contributes
$2e^{2}/h$ to the conductance. An in-plane magnetic field
$B_{\parallel}$ was applied parallel to the direction of transport
in the QPC to induce Zeeman splitting. At $B_{\parallel}=10$ T, the
spin-degeneracy is lifted (red line in Fig.~\ref{fig:1}(b)) and the
conductance is quantized in units $e^2/h$. The evolution of the spin
splitting with applied magnetic field is shown in
Fig.~\ref{fig:1}(a), where the transconductance $dG/dV_{g}$ is
plotted as a function of the gate voltage and magnetic field. The
light regions mark the 1D subband edges, corresponding to the rises
between conductance plateaux where $dG/dV_{g}$ is maximum. The
spin-splitting $\Delta E_{z}$ increases with applied magnetic field,
which is observed as an increasingly large splitting $\delta V_{g}$
between the bright regions in Fig.~\ref{fig:1}(a).

Unfortunately, the splitting $\Delta E_{z}$ cannot be obtained
directly from Fig.~\ref{fig:1}(a), which only gives $\delta V_{g}$.
To calculate the $g$-factor, it is necessary to convert $\delta
V_{g}$ into the subband energy scale, which is achieved by measuring the
splitting in gate voltage due to an applied d.c.~source-drain bias
$V_{sd}$. We do this using the method developed by Patel \emph{et
al.}~\cite{patel91a} that combines measurements of the splitting due
to the field $\delta V_{g}/\delta B_{\parallel}$
(Fig.~\ref{fig:1}(a)) with measurements of the splitting due the
source-drain bias $\delta V_{g}/\delta V_{sd}$ to give the absolute
value of the $g$-factor:
\begin{eqnarray}
\label{eqn:1} |g^{*}|=\frac{1}{\mu_{B}}\frac{d(\Delta
E_z)}{dB_{\parallel}}=\frac{1}{\mu_{B}}\frac{d(\Delta
E_z)}{dV_{g}}\frac{dV_{g}}{dB_{\parallel}}=\frac{e}{\mu_{B}}\frac{\delta
V_{sd}}{\delta V_{g}}\frac{\delta V_{g}}{\delta B_{\parallel}}
\end{eqnarray}

\begin{figure}
\includegraphics[width=8.5cm]{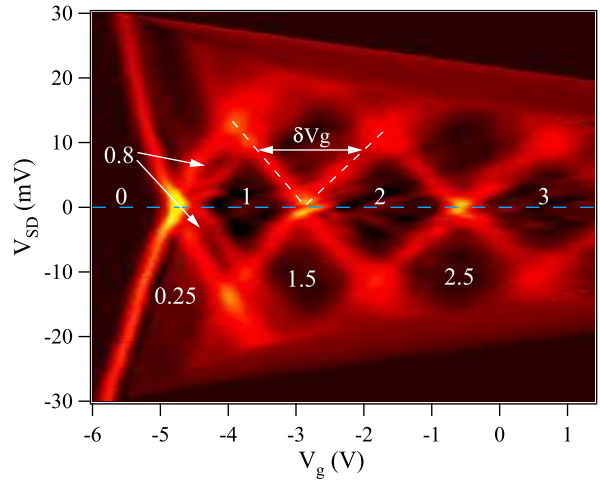}
\caption{\label{fig:2} (Color online) Transconductance $dG/dV_{g}$
plotted as a function of gate voltage $V_{g}$ (horizontal axis) and
source-drain bias $V_{sd}$ (vertical axis) at $B_{\parallel}=0$.
Regions of yellow and red correspond to a large amplitude of
$dG/dV_{g}$, indicating the locations of subband transitions. The
conductance $G$ is labeled on each plateau in units of $2e^{2}/h$.
Arrows indicate the $G\simeq0.8(2e^{2}/h)$ plateaux that are linked
to the 0.7 feature.~\cite{patel91b,kristensen00} The blue dashed
line indicates the region of the plot that is equivalent to the blue
trace in Fig.~\ref{fig:1}(b).}
\end{figure}

The source-drain bias measurements are shown in Fig.~\ref{fig:2},
where the transconductance $dG/dV_{g}$ is plotted as a function of
$V_{g}$ and $V_{sd}$ at $B_{\parallel}=0$ (data have been corrected
for the d.c.~bias dropped across the series resistance). Similar to
Fig.~\ref{fig:1}(a), the light regions correspond to the transitions
between 1D subbands (large $dG/dV_{g}$), marking the 1D subband
edges. As the applied bias $V_{sd}$ is increased, the subband
transitions (light regions) split in gate voltage by $\delta V_{g}$,
and conductance plateaux at half-integer $2e^{2}/h$ appear between
the split transitions.~\cite{patel91b,kristensen00} Since the
splitting $\delta V_{g}$ is directly proportional to $eV_{sd}$, the
energy scales associated with $\delta V_{g}$ in Fig.~\ref{fig:2} are
obtained by calculating $e\delta V_{sd}/\delta
V_{g}$.~\cite{patel91b,kristensen00}

The transconductance maxima in Fig.~\ref{fig:2} cross when the
applied source-drain bias is equal to the subband spacing: $\Delta
E_{n,n+1}=eV_{sd}$,~\cite{patel91b,kristensen00} allowing the
subband spacing to be directly extracted from Fig.~\ref{fig:2}.
Values of $\Delta E_{n,n+1}$ calculated from these crossing points are listed in
Table~\ref{tab:table1} and exceed 10 meV.  The subband spacings are 
consistent with those obtained using magnetic depopulation 
measurements,~\cite{martin08b} and are double the spacings measured for
etched In$_{0.47}$Ga$_{0.53}$As quantum wires with InAlAs 
barriers.~\cite{sugaya01} In Refs.~16 and~17 the `0.7
feature' evolves into plateaux with $G\simeq0.85(2e^{2}/h)$ at
finite source-drain bias. In Fig.~\ref{fig:2}, we observe similar
`shoulder' plateaux with $G\simeq0.8(2e^{2}/h)$, despite there being
no clear 0.7 feature at $V_{sd}=0$. Oscillatory structure is also
observed on the conductance plateaux in Fig.~\ref{fig:2}, which has
recently been linked to standing waves in the QPC~\cite{lindelof08}
and will be discussed elsewhere.

\begin{figure}
\includegraphics[width=9cm]{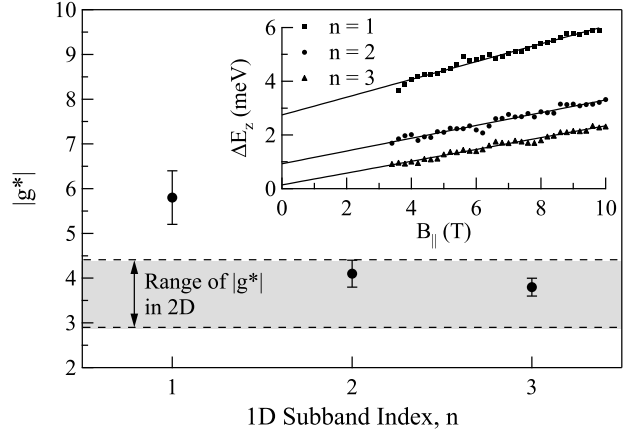}
\caption{\label{fig:3} Effective Land\'e \emph{g}-factor $|g^{*}|$ in the QPC
as a function of 1D subband index $n$. Dashed lines bound the range
of $|g^{*}|$ previously measured for 2D Ga$_{x}$In$_{1-x}$As
systems.~\cite{vehse86,dobers89,savelev96,schapers07} Inset: Zeeman 
splitting $\Delta E_{z}$ plotted vs magnetic field $B_{\parallel}$. Linear 
fits are calculated between 3~T~$<B_{\parallel}\le10$~T and are
then extrapolated to $B_{\parallel}=0$.}
\end{figure}

The Zeeman energies $\Delta E_{z}$ and $g$-factors are calculated using
Equation~(\ref{eqn:1}) and the data extracted from Figs.~\ref{fig:1} 
and~\ref{fig:2}. The splitting rates $\delta V_{g}/\delta B_{\parallel}$ and
$e\delta V_{sd}/\delta V_{g}$ are listed 
in Table~\ref{tab:table1}.
The Zeeman splitting of the 1D subbands $\Delta E_{z}$ 
is plotted against $B_{\parallel}$ in the 
inset of Fig.~\ref{fig:3}, demonstrating 
a linear relationship for $B_{\parallel}\gtrsim3$~T
where the subband transitions are clearly resolved. Although the linear fits for
the first two subbands do not extrapolate to zero at $B_{\parallel}=0$ 
(due to the presence of the 0.7 feature and its analog
at 1.7), still the $g$-factor can be extracted for each 
subband using Eq.~(\ref{eqn:1}).~\cite{patel91a,thomas96} The values of $|g^{*}|$  
are plotted as a function of $n$ in Fig.~\ref{fig:3}. For the $n=2$ and
$n=3$ subbands, we find that $|g^{*}|$ falls within the range
previously reported for 2D Ga$_{x}$In$_{1-x}$As
systems.~\cite{vehse86,dobers89,savelev96,schapers07} However, in
contrast with Ref.~8, we obtain
$|g^{*}|=5.8\pm0.6$ for $n=1$, confirming the presence of a 1D
enhancement in our Ga$_{0.25}$In$_{0.75}$As QPC.

In Ref.~8, the Zeeman splitting was measured with fields 
3~T~$\le B_{\perp}\le8$~T applied perpendicular, rather than parallel, to the quantum well. 
This produces magnetic confinement due to Landau quantization that adds to the QPC's
electrostatic confinement. Ref.~8 reports 1D subband spacings that are smaller than  
the Landau level spacing for these fields.~\cite{schapers07} This is a non-physical 
result, since the addition of 1D confinement to a 2D system in a perpendicular field  
cannot \emph{decrease} the energy level spacing.  For example,  
the 1D subband spacing is reported to be only 8.4 meV at $B_{\perp}=4$~T when the 2D  
Landau level spacing is already 11.9~meV.~\cite{schapers07} Thus we suggest that the value of 
$|g^{*}|$ reported in Ref.~8 has been underestimated, which 
explains why no enhancement was observed. In contrast, our data are obtained 
with an in-plane magnetic field in the absence of Landau quantization. 
Our measurements show an enhancement of the $g$-factor as the QPC becomes 
more one-dimensional in agreement with previous results in 
1D GaAs systems.~\cite{patel91a,thomas96,daneshvar97,danneau06}

In summary, we have demonstrated that 1D confinement in a narrow-gap
semiconductor such as GaInAs can result in a significant enhancement
of the Zeeman splitting.  Although the enhancement in $|g^{*}|$ is
consistent with the trend observed in GaAs
QPCs,~\cite{patel91a,thomas96,daneshvar97,danneau06} the magnitude of $|g^{*}|$ measured
for GaInAs QPCs is six times greater. Additionally, the subband
spacing in our devices is quite large.~\cite{martin08b} Combined,
the large $g$-factors and subband spacings found in etched GaInAs
QPCs make them suitable for applications requiring spin-sensitivity
at higher temperatures and with small magnetic fields.

We acknowledge financial support by the Australian Research Council,
the National Science Foundation, and the Research Corporation.

\clearpage


\clearpage

\begin{table*}
\caption{\label{table1} Subband-dependent parameters of the QPC: subband spacing
$\Delta E_{n,n+1}$, subband splitting rates $\delta V_{g}/\delta B_{\parallel}$
and $e\delta V_{sd}/\delta V_{g}$, and effective Land\'e $g$-factor $|g^{*}|$.}
\begin{tabular*}{0.7\textwidth}{@{\extracolsep{\fill}} c c c c c }
\hline \hline
Parameter & & $n=1$ & $n=2$ & $n=3$ \\
\hline
$\Delta E_{n,n+1}$ & (meV) & $13.1\pm0.5$ & $11.1\pm0.5$ & $9.9\pm0.5$\\
$\delta V_{g}/\delta B_{\parallel}$ & (V/T) & $0.029\pm0.003$ & $0.0371\pm0.0015$ & $0.0500\pm0.0015$\\
$e\delta V_{sd}/\delta V_{g}$ & (meV/V) & $11.6\pm0.6$ & $6.4\pm0.3$ & $4.4\pm0.2$\\
$|g^{*}|$ & - - & $5.8\pm0.6$ & $4.1\pm0.3$ & $3.8\pm0.2$\\
\hline\hline
\end{tabular*}
\label{tab:table1}
\end{table*}

\clearpage

\clearpage

\clearpage

\end{document}